# Monolithic 3D Integration for Null Convention Logic (NCL)-Based Asynchronous Circuits

Xiameng Zhang[1], Kushal K. Ponugoti[2], Ashiq A. Sakib[3], Madhava Sarma Vemuri[1]
[1]School of STEM, University of Washington Bothell, WA, USA 98011
[2]Department of Electrical and Computer Engineering, North Dakota State University, ND, USA 58105
[3]Department of Electrical and Computer Engineering, Southern Illinois University Edwardsville, IL, USA 62026,
Email: xiameng@uw.edu, kushal.ponugoti@ndsu.edu, asakib@siue.edu, madhava@uw.edu

***Abstract*—As the demand for high-speed and low-power electronics continues to grow, the quasi-delay-insensitive (QDI) asynchronous domain of digital design has emerged as a promising alternative to traditional clock-based designs. However, the adoption of the paradigm has been greatly limited due to the lack of mature computer-aided design (CAD) tools and a substantially larger area footprint, owing to various architectural constraints. Monolithic-3D (M3D) technology has recently paved the way for manufacturing highly dense integrated circuits (ICs) through sequential integration, resulting in reduced area footprint, shorter wirelengths, and increased performance. In this study, we integrate M3D technology with QDI Null Convention Logic (NCL) and propose a design methodology for the implementation of M3D-based NCL standard cells, aimed at mitigating the area inefficiencies of traditional planar or 2D counterparts. Furthermore, we employed the threshold gates to design an M3D-NCL unsigned array multiplier circuit. Simulation results suggest that, for a conservative wirelength reduction resulting from M3D implementation, a substantial area reduction of 44% can be achieved while simultaneously reducing delay and power by ∼31%, and ∼17%, respectively.**

***Index Terms*—QDI Asynchronous Circuits, Null Convention Logic, Vertical Integration, Monolithic 3D ICs**

## I. Introduction

The demand for energy-efficient and high-performance integrated circuits (ICs) is growing rapidly. For decades, the synchronous domain of digital IC design has been the standard, effectively accommodating this ever-increasing demand. However, the clock-based design style is rapidly approaching its limits in recent years as device dimensions continue to shrink and designs become exceedingly complex. Clock management, clock distribution, increased power dissipation, susceptibility to process, voltage, temperature (PVT) variations, and timing inconsistencies surface as significant challenges that are becoming substantially difficult to resolve. Consequently, there is a growing interest in exploring alternatives to synchronous design methodologies to circumvent the aforementioned challenges. Quasi-delay-insensitive (QDI) asynchronous design offers a fundamentally transformative approach, eliminating the global clock and employing handshake-driven synchronization, making it resilient to timing inconsistencies and inherently 'correct by construction' [1]. Null Convention Logic (NCL) is a mature and widely adopted QDI paradigm, as NCL circuits exhibit low power performance, generate less noise and electromagnetic interference (EMI), are robust against PVT variations, and better handle operating conditions [2], [3]. Moreover, the formal foundations of QDI design allow the use of robust formal verification methodologies to prove correctness and handshake integrity, strengthening resistance against malicious modifications and subtle design flaws. This combination of intrinsic robustness and formal verifiability makes NCL particularly attractive for trustworthy hardware design [4], [5].

Another major challenge facing today's semiconductor industry is the slowing down of the transistor's physical scaling, attributed to the physical and manufacturing limits of conventional two-dimensional (2D) technology. Innovative physical design techniques are being investigated to address the constraints of planar 2D ICs. The three-dimensional (3D) integration allows multiple layers of devices to be vertically integrated, leading to shorter wirelengths, reduced routing congestion, and higher functional density [6]. This makes it a promising method to improve performance and efficiency in modern chips in the coming years. Among the various 3D integration approaches, die stacking facilitates the placement of traditional 2D dies atop the interposer substrate via μ-bumps and Through-Silicon-Via (TSV) technology. TSVs are known to occupy a significant area, leading to an increase in area overhead [7]. Recently, Monolithic-3D (M3D) technology has emerged as a more viable alternative, paving the way for manufacturing highly dense ICs through sequential integration, resulting in reduced footprint area, shorter wirelengths, and increased performance. Unlike TSV-based methods, which require large vertical interconnects, M3D uses fine-grained metal interlayer vias (MIVs) [8].

The research presented herein aims to study the application of M3D technology in NCL circuits. The larger area footprint is one of the primary factors that has impeded the widespread adoption of NCL circuits, despite their commercial success [9]. The integration of M3D can offset the larger area requirements of NCL, thereby introducing new classes of applications that can benefit from robust, energy-efficient, delay-insensitive asynchronous computing without sacrificing an excessive amount of silicon real estate. With this motivation, the following are the major contributions of this work.

1) We propose the first-ever design and analysis methodology for M3D-based implementations of static NCL (M3D-NCL) threshold gates.

2) We evaluate the power, performance, and area (PPA) metrics of M3D-NCL standard gates and compare those to their corresponding 2D implementations.
3) We employ the M3D-NCL gates to design an unsigned NCL array multiplier to study the benefits of M3D implementation in comparison to conventional 2D implementation for large-scale circuits.
4) Simulation results for the NCL gates suggest that, for a conservative wirelength reduction resulting from M3D implementation, a substantial area reduction of 44% on average can be achieved while simultaneously reducing the average delay, skew, and power of the NCL gates by 10.5%, 10.2%, and 10.3%, respectively.
5) Simulation results for the multiplier design suggest that the M3D version outperformed the 2D version in all categories while utilizing half the area.

The organization of the rest of the paper is as follows: Section II provides the background and related works of the current study. Section III discusses the proposed methodology to design and analyze the M3D implementation of NCL gates. Section IV presents the simulation results and discusses the systematic study conducted in the paper. Finally, the concluding remarks are presented in Section V.

## II. Background and Related Works

### A. NCL Background

NCL has several unique features that separate it from its synchronous/Boolean counterparts. Unlike Boolean logic, NCL employs a 1-hot encoding scheme (1-of-n), most commonly the dual-rail (DR) logic (1-of-2), to encode data and eliminate timing references. In DR logic, two wires (or rails) are utilized to represent one bit of information, as opposed to a single wire in Boolean logic. A DR variable, $D$, can assume one of the three permissible values from the set $\{$DATA0, DATA1, NULL$\}$, where DATA0 ($D^1D^0 = 01$) and DATA1 ($D^1D^0 = 10$) correspond to Boolean logics 0 and 1, respectively. ($D^1D^0 = 00$) is a spacer state, commonly referred to as the NULL state, which indicates the unavailability of the data. ($D^1D^0 = 11$) constitutes an invalid state. The NCL framework resembles a traditional synchronous framework, with each stage consisting of an input and output NCL registration unit and one combinational logic (C/L) in between, as illustrated in Fig. 1a. The input wavefronts are, however, controlled by local handshaking signals, rather than a global clock signal. A 4-phase handshaking protocol is employed, in which the completion detection unit (C/D) and registers maintain an alternating sequence of NULL and DATA to differentiate between two unique data sets [2], as shown in Fig. 1(b).

The NCL logic, including registers and completion, comprises 27 fundamental gates with *hysteresis*, i.e., state holding capability. Each gate is characterized by unique input configurations and threshold levels. Together, these gates can implement any function of up to 4 variables. The threshold gates are classified into two categories: non-weighted and weighted. A non-weighted gate, also known as an *m-of-n* gate, is denoted as *THmn*, where $n$ is the number of inputs and $m$ ($1 \leq m \leq n$) is the threshold (the minimum number of inputs that must be asserted to assert the output) (Fig. 1c). A weighted gate is represented as $THmnWw_1w_2...w_R$, where $w_R$ ($1 < w_R \leq m$) is the weight corresponding to input $R$ (Fig. 1d). Fig. 1e illustrates the generic CMOS template for the static implementation of any NCL threshold gate. The template consists of four main blocks: *SET*, *RESET*, *Hold0*, and *Hold1*. The gate threshold function is implemented by the *SET* block, while the *RESET* block deasserts the output when all inputs are deasserted. The *Hold* blocks implement hysteresis functions that retain the output when neither *SET* nor *RESET* is active, utilizing two feedback transistors, as indicated in Fig. 1e. NCL gates can also be implemented in a semi-static manner, in which the *Hold* blocks are eliminated and replaced by a weak-feedback inverter arrangement [2].

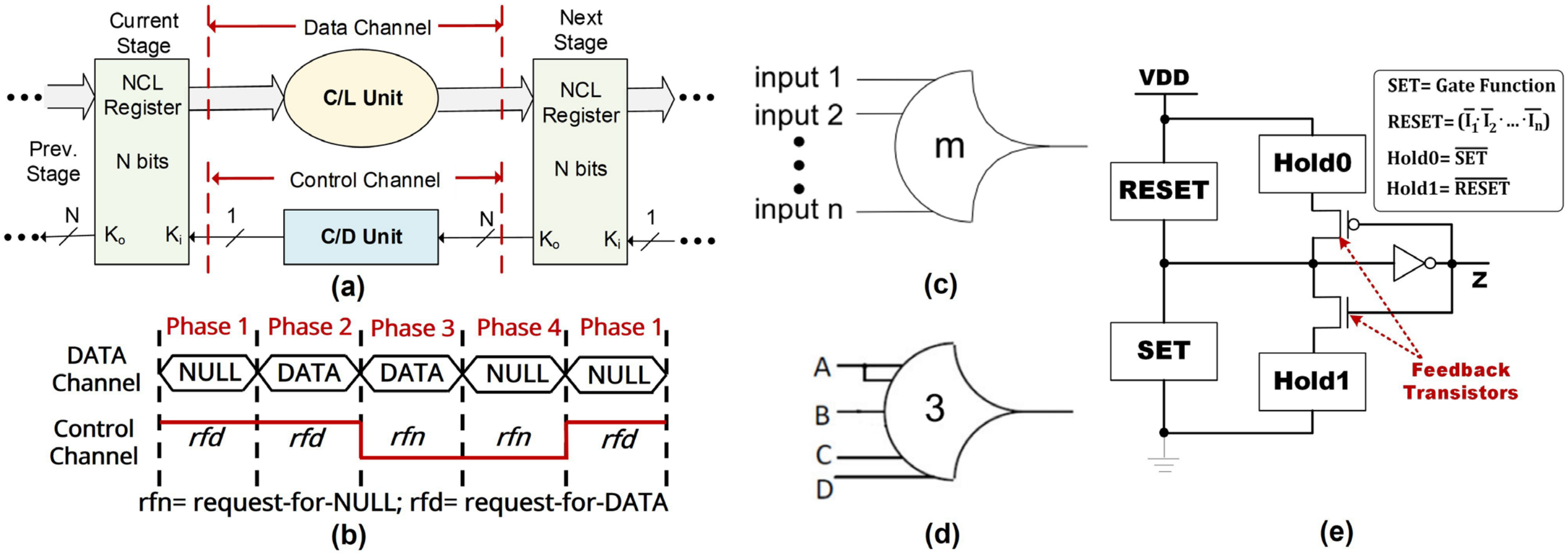

Fig. 1: (a) NCL framework, (b) 4-phased handshaking scheme, (c) Generic non-weighted gate symbol, (d) TH34w2 weighted gate, and (e) CMOS static implementation template for NCL gates.

While this decreases the transistor count as compared to the static version, it does not inherently result in a reduced silicon area. This is because pmos transistors require substantial sizing up to maintain the necessary driving strength for proper functionality. Furthermore, semi-static gates exhibit reduced speed and higher energy consumption per switching relative to their static counterparts [9]. Therefore, the research presented herein primarily focuses on the static implementations of NCL gates.

### B. M3D Implementation Styles

M3D offers various abstract design styles, including block-level, gate-level, and transistor-level, as depicted in Fig. 2 [10]–[12]. In block-level design style, the conventional 2D IP blocks are partitioned into different layers, as depicted in Fig. 2a. Compared to other implementations, this style offers the least area savings, as it does not allow finer integration of Metal Inter-layer Vias (MIVs) [13]. In gate-level design style, as depicted in Fig. 2b, the gates are placed in multiple layers, and MIVs connect the input and output pins in different layers. This approach allows the utilization of existing libraries and has reported area savings of up to 50% [12]. However, it does not address the reliability concerns that result from sequential integration [14]–[16]. In a transistor-level design style, NMOS and PMOS are assigned to different layers, requiring only local interconnects to route the bottom layers, as depicted in Fig. 2c. This results in fewer manufacturing steps, resulting in controlling variations arising from M3D integration [17]. The role of MIV is to facilitate interconnections to active and gate regions of various devices in multiple layers. [13] indicates that transistor-level implementations can result in a 12% to 40% reduction in area. The theoretical 50% reduction of area is not observed due to a mismatch between transistor devices in different layers and additional area overhead caused by the presence of MIVs in the top layer. Additionally, a Keep-Out Zone (KOZ) around the MIV is a critical requirement to prevent it from biasing the adjacent devices in the top layer, which further reduces the area savings [18], [19].

### C. Related Works

NCL circuits must follow delay-insensitive (DI) constraints, such as *input-completeness* and *observability*[1], and maintain handshaking protocols for synchronization, leading to significant area overhead compared to their clocked counterparts. Over the years, many research efforts have focused primarily on NCL implementations with fewer constraints to enhance area efficiency. In [20], a modified NCL architecture, NCL_X, was proposed that allows the C/L unit to be relaxed, i.e., input-incomplete and unobservable, substantially reducing the area. However, the DI constraints are preserved by additional completion logic, resulting in a handshaking network that is both larger and slower. Multiple relaxation schemes were proposed in [21] and [22], aiming to achieve an optimal balance between strict and relaxed implementation. [23] is a more recent work that intelligently combines [20] and [21] and expands upon both to yield smaller and faster NCL circuits. At the circuit level, the minimization of the fundamental threshold gates' area was also the focus of numerous prior studies, both in CMOS [24], [25], [26] and beyond CMOS technologies [27], [28], [29]. However, those works relied on traditional 2D technology, limiting the potential for substantial area optimization. There exist very few works that have explored the potential of 3D integration in the context of clockless design. In [30], researchers have focused on the use of vertical integration techniques to design QDI asynchronous circuits using Through-silicon-via (TSV) based integration. However, TSVs are significantly larger than MIVs (> 200×) and require large KOZs to limit the degradation of devices around TSVs [31]. Meanwhile, [32] introduces a gate-level M3D implementation methodology for desynchronized circuits and reports reductions in wirelength and area of up to 35% and 50%, respectively. However, as mentioned earlier, the transistor-level M3D integration provides better reliability than gate-level integration, and to the best of our knowledge, no prior research has been reported on the transistor-level M3D implementation for QDI NCL asynchronous circuits.

[1]Input completeness requires that no output of a C/L circuit can transition from Null-to-Data (or Data-to-Null) until all inputs have transitioned from Null-to-Data (or Data-to-Null). Observability requires that every switching gate must be observable at the output; i.e., switch at least one of the primary outputs.

## III. Proposed Methodology

### A. Process and Design Considerations

Since there are no open-source process design kits (PDKs) for the M3D implementation of asynchronous logic, we have adapted the NCSU FreePDK45 model library and made realistic assumptions when designing M3D-IC for the Bulk-Si process [33]. In the conventional planar implementation, both the PMOS and NMOS transistors are fabricated on the same layer, as shown in Fig. 3(a). Silicon trench isolation (STI) limits the interactions of different PMOS and NMOS devices. For the proposed M3D implementation, we have considered a two-tier transistor-level abstract style with NMOS and PMOS transistors in the top and bottom layers, respectively, as shown in Fig. 3(b) [34]. The interlayer dielectric (ILD) limits electrical coupling and improves thermal conductivity between layers [35]. The M3D follows a sequential integration process (i.e., the bottom, ILD, and top layers are created sequentially). Since the top layers are created on the fabricated bottom layer, a thermal budget (< 500ºC) is set to limit the degradation of transistor devices in the bottom layers [14]–[16]. Metal routing layers are created inside the interconnect dielectric (ID). MIVs connect the metal interconnects in both layers to facilitate interconnections between the top and bottom layers. An oxide liner is created to prevent the MIV from biasing the devices in the top layer. An additional keep-out-zone (KOZ) is also required for the Bulk-Si around MIV to limit the interactions of MIV and neighboring devices [36].

The process and design parameters considered for this work are presented in Table I. The nominal values are extracted from the NCSU FreePDK45nm library [33]. We considered the length of the channel region to be 50 nm, and the length and width of the source and drain regions to be 90 nm. The width and pitch of the interconnect wires are 65 nm and 130 nm, respectively. We have considered the copper connecting region to have a sheet resistance of 0.38Ω/sq and a capacitance per unit length of 179.93 fF/mm, based on [33], [37]. The metal vias connect 1) the interconnect to active regions (M1 - source/drain) and 2) the interconnect to the gate region (M1 - gate). We have considered the via resistance to be 6Ω, and the MIV resistance and capacitance to be 5.5Ω and 0.04 fF, respectively, which is consistent with [13]. The MIV thickness and the KOZ around the MIV are both considered to be 50nm, as per [36].

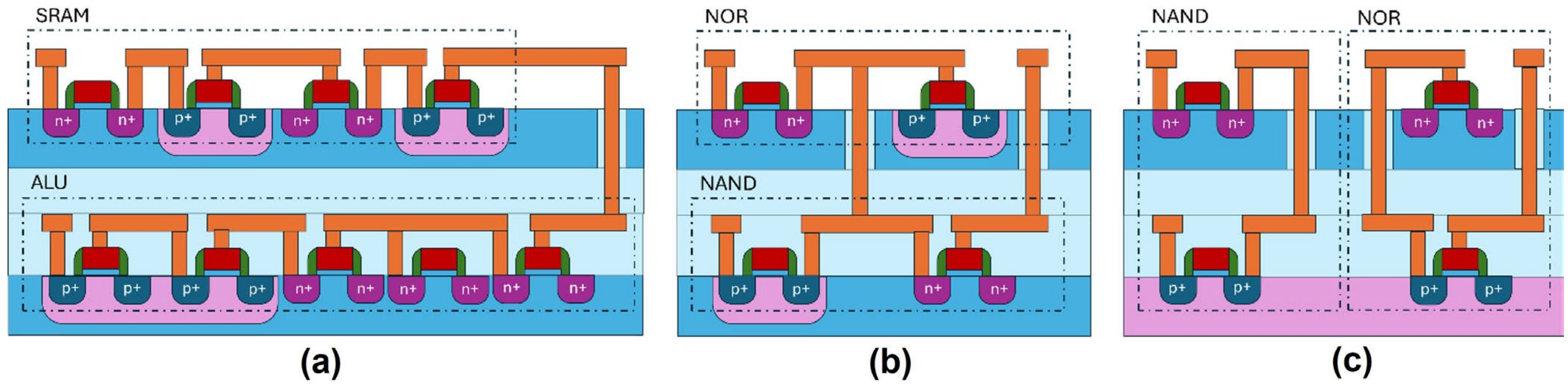

Fig. 2: M3D implementation styles: (a) block-level, (b) gate-level, and (c) transistor-level.

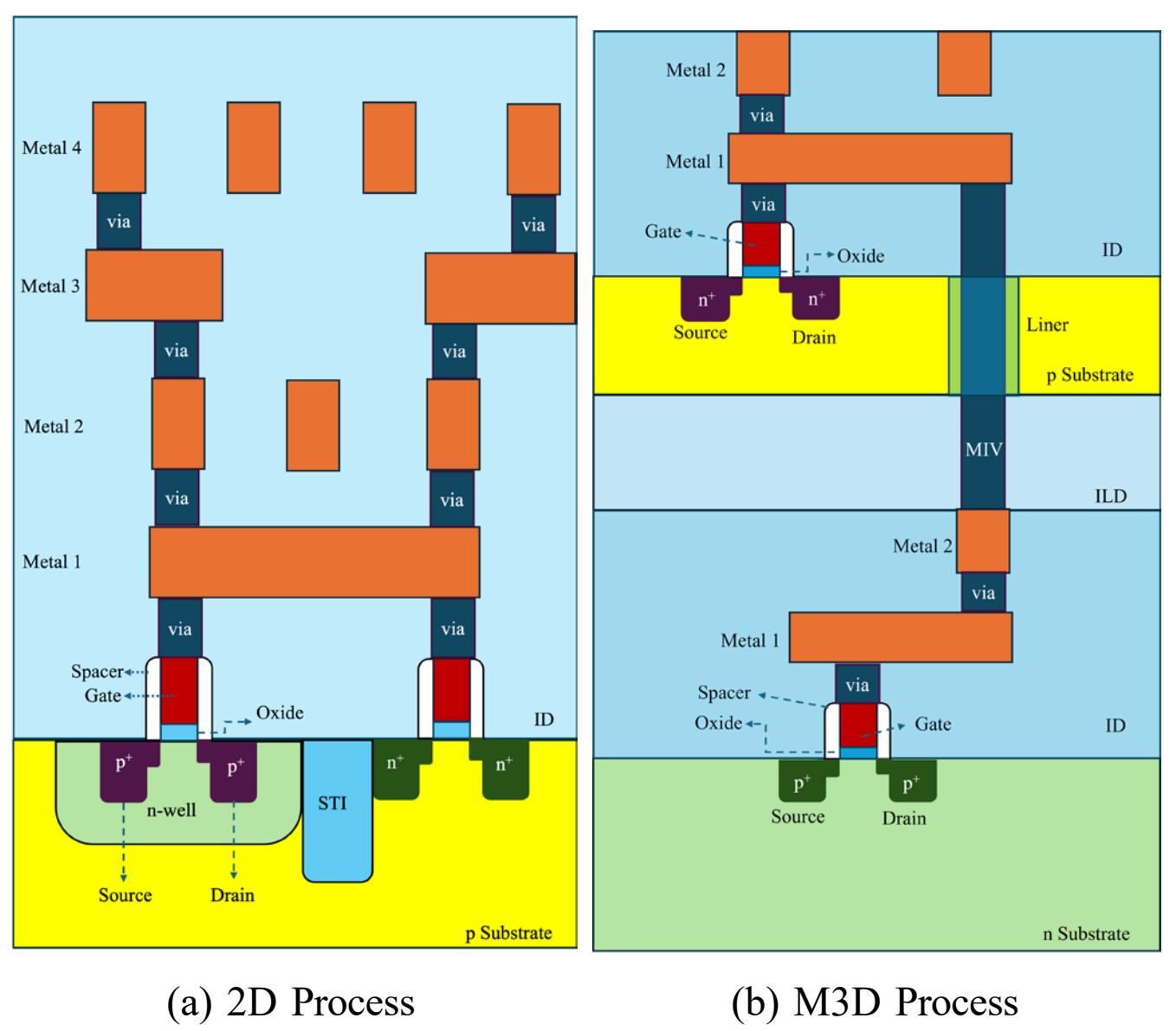

Fig. 3: 2D and proposed M3D process comparison (not to scale)

### *B. Overview of Selected NCL Threshold Gates*

As mentioned in Section IIA, the 27 NCL gates can implement any function of up to 4 variables. In this work, we focus on 6 specific gates: TH22 ($PMOS$ : 6, $NMOS$ : 6), TH24 ($PMOS$ : 13, $NMOS$ : 13), TH34 ($PMOS$ : 13, $NMOS$ : 11), TH24comp ($PMOS$ : 9, $NMOS$ : 9), THand0 ($PMOS$ : 10, $NMOS$ : 10), and TH54w322 ($PMOS$ : 11, $NMOS$ : 10). These gates are chosen as they cover both weighted and non-weighted gates of varying complexities and transistor counts. By analyzing this set, we can better understand how gate complexity affects performance and integration in our proposed M3D design. Each gate has two to four inputs ($a$, $b$, $c$, and $d$) and one output ($Z$). The TH22 gate is the simplest gate in the set that behaves like a 2-input Muller C-element [38], asserting/deasserting the output only when both inputs are asserted/deasserted; otherwise, holding the previous value. Based on the threshold, the *SET* function for this gate can be expressed as $Z = ab$. The TH24 gate, with four inputs and a threshold of two, is the largest NCL gate, which gets asserted when at least two of its inputs are asserted, making the *SET* expression to be $Z = ab + ac + ad + bc + bd + cd$. Similarly, the TH34 gate with a threshold of three has the following *SET* function: $Z = abc + abd + acd + bcd$. The *SET* functions of the TH24comp and THand0 gates are $Z = ac + ad + bc + bd$ and $Z = ab + bc + ad$, respectively. The weighted gate in our study is the TH54w322 gate. The gate has four inputs, with the first input $a$ being assigned a weight of three, inputs $b$ and $c$ each being assigned a weight of two, and the remaining input $d$ having a weight of one. The output is asserted when either $a$ and $b$ are asserted, $a$ and $c$ are asserted, or when all of $b, c$ and $d$ are asserted, satisfying the threshold of five (i.e., $Z = ab + ac + bcd$).

TABLE I: Process and design parameters used in the study for 2D and M3D implementation

| Parameter | Description | Value |
|---|---|---|
| $L_G$ | Length of Channel | 50 nm |
| $l_{src}$ | Length of source/drain region | 50 nm |
| $w_{src}$ | Width of source/drain region | 90 nm |
| $t_{ILD}$ | Interlayer dielectric thickness | 120 nm |
| $t_{miv}$ | MIV thickness | 50 nm |
| $w_{M1}$ | Width of M1 interconnect | 65 nm |
| $w_{src}$ | Pitch of M1 Iinterconnect | 130 nm |
| $w_{gate}$ | Width of gate region | 50 nm |
| $R_{int,sq}$ | Sheet resistance of interconnect | 0.38 Ω/sq |
| $R_{via}$ | Via Resistance | 6 Ω |
| $C_{int}$ | Interconnect capacitance per unit length | 179.93 fF/mm |
| $R_{MIV}$ | MIV resistance | 5.5 Ω |
| $C_{MIV}$ | MIV capacitance | 0.04 fF |

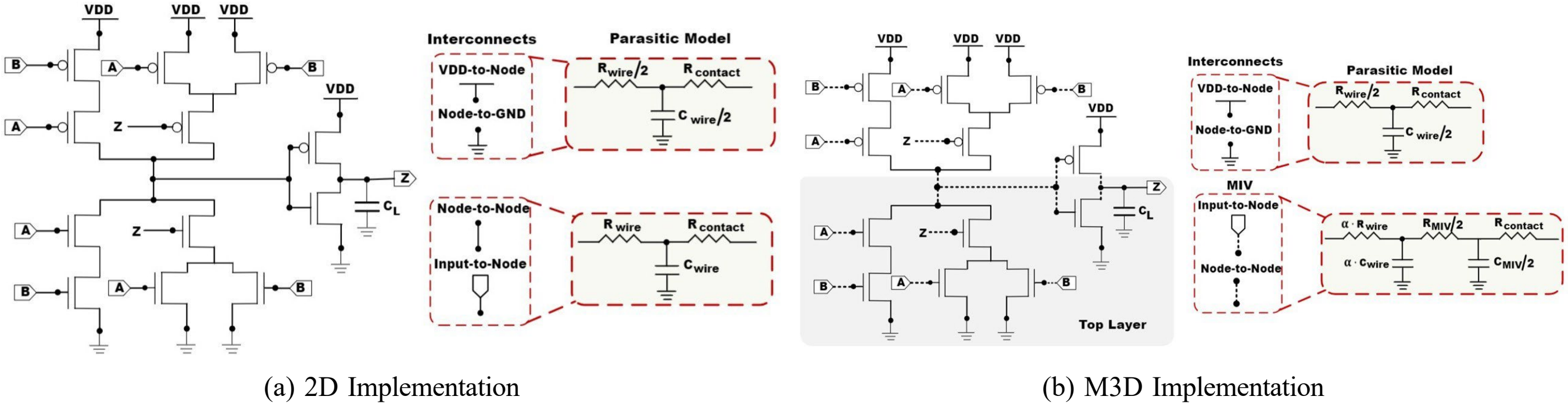

(a) 2D Implementation (b) M3D Implementation

Fig. 4: 2D and M3D implementation of TH22 circuit schematic with the RC prasitic model

### C. Parasitic Model Used in the Study

In this section, we have used the standard TH22 gate to explain the parasitic model of conventional 2D and the proposed M3D implementation. For both cases, there are 4 common metal routing scenarios, such as: 1) VDD-to-Node, 2) Node–to-GND, 3) Node-to-Node, 4) Input-to-Node as shown in Fig. 4. The VDD-to-Node and Node-to-GND connection scenario interconnects power rails such as VDD and GND to the transistor devices' active regions (i.e., the source and drain). The Node-to-Node connects the active regions of PMOS and NMOS devices. The Input-to-Node connects the inputs, such as $a$, $b$, $c$, and $d$, to the gate regions of the devices. RC parasitics, such as $R_{wire}$ and $C_{wire}$, for the interconnect regions are calculated using the standard cell height. The wires are considered to be extended for half the height of the cell in scenarios that connect to the power and ground rails [39]. In the M3D implementation, since the PMOS and NMOS devices are placed in different layers, MIV is used to connect the interconnect regions in different layers. Subsequently, the associated RC parasitics, such as $R_{MIV}$ and $C_{MIV}$ are added into the parasitic model of the M3D implementation, as shown in Fig. 4(b). The scaling factor $\alpha$ is used to characterize the reduction of RC parasitics from metal interconnect routing, which is a consequence of the improved routing congestion caused by the reduced cell height in both layers. *It is important to note that we have associated the scaling factor to interconnect scenarios such as Node-to-Node and Input-to-Node, as these two are more substantially impacted by M3D implementation than other interconnect scenarios*. Interconnect capacitances, such as fringing capacitance and metal coupling capacitance, are ignored to limit the complexity of the design.

## IV. Simulation Results and Performance Assessment

### A. Systematic Study of PPA Metrics of NCL Standard Cells

Based on the parasitic model described in the previous section, we designed both 2D and M3D NCL standard cells, as outlined in Section III-B, utilizing a commercial SPICE circuit simulator. Custom cells with a height of 14 tracks were designed using the model libraries provided by the NCSU FreePDK45nm. A 1 fF load capacitance has been considered for each gate. A systematic variation in wirelength reduction, ranging from 20% to 40%, was conducted to examine the effects of M3D integration. This variation is represented by the parameter $\alpha$, which quantifies the extent of physical reduction due to vertical stacking. Specifically, the $\alpha$ values of 0.8, 0.7, and 0.6 correspond to the wirelength reductions of 20%, 30%, and 40%, respectively. These values were chosen to capture both conservative and aggressive levels of 3D integration and to assess how different degrees of vertical compaction affect key performance metrics. Table II reports the power, performance, and area (PPA) metrics when the wire length reduction is 30%, i.e., ($\alpha = 0.7$). These metrics were chosen because they collectively represent the temporal and dynamic behavior of the test circuits. Delay and skew influence timing robustness and throughput, while power is critical for energy efficiency and thermal reliability, especially in vertically stacked architectures.

The simulation results indicate that the propagation delay ($T_D$) is reduced by approximately 11% on average in the M3D implementations relative to the 2D counterparts, with the TH22 gate exhibiting the most notable improvement of around 12%. The average output skew ($T_S$) was reduced by 10.2% on average for the M3D case. The average power consumption of the gates was lowered by around 10% on average. The TH22 and TH34 gates have shown the highest power savings, reaching up to 11%, in comparison to others. The total area footprint of the 2D implementations is compared with that of the M3D implementations, accounting for the additional MIV area overhead for the Bulk-Si process. The results indicate a 44% decrease in average area utilization, with TH24 exhibiting the maximum reduction of 46%.

Figure 5 compares the PPA metrics of the different NCL gates across varying scaling factors ($\alpha$). We observe that the highest performance gains are achieved when the wirelength is reduced by 40% ($\alpha = 0.6$). In this configuration, the benefits of shorter interconnects and reduced parasitics are most pronounced. The delay and skew improve by up to 16% and 18%, respectively, while the power consumption shows a reduction of up to 16% relative to the baseline 2D implementation. These improvements come from placing logic elements closer together in the M3D layout, which shortens the distance signals need to travel and reduces the energy during switching.

In contrast, when the wirelength reduction is limited to 20% ($\alpha = 0.8$), the overall improvements are modest. All the performance parameters, i.e., delay, skew, and power, are improved by 6% on average. These results suggest that without compromising the area benefits, limited reductions in wirelengths may result in reduced power and performance benefits arising from M3D.

### *B. Performance Evaluation for Large-Scale NCL Circuits*

The goal of this section is to present the impact of M3D integration on large-scale NCL circuits and provide a comparative analysis of performance against their 2D counterparts. A QDI 4×4 NCL unsigned multiplier was synthesized at the gate level for this purpose. A Boolean RTL netlist was first generated using Yosys [40], an open-source RTL synthesizer. The generated Boolean netlist was then converted to an NCL gate-level netlist by the UNCLE synthesis tool, using the tool's intermediate output before the addition of registration and completion logic [41]. The multiplier created utilized multiple threshold gates, including the TH22, TH24comp, and THand0 NCL gates, from Table II. The circuit was then converted to SPICE models using the threshold gate implementations based on Section III. The C/L of the 4x4 NCL multiplier requires a total of 2124 transistors.

We evaluate the performance of the M3D-NCL multiplier in terms of area, power, and delay, and compare it with that of its 2D counterpart. The M3D-NCL multiplier demonstrates a 44.5% reduction in area utilization compared to the 2D version. We report the average power and delay of the designed multipliers using multiple sets of random input combinations. For the nominal case ($\alpha = 0.7$), the 2D and M3D versions consumed 56µW and 46.7µW, respectively, indicating an improvement of ∼17%. The 2D-NCL multiplier demonstrates a worst-case delay of 1.98ns, while the M3D-NCL multiplier has a delay of 1.37ns, demonstrating an improvement of 30.8%. Fig. 6 also presents the % improvement in power and delay for varying scaling factors. Finally, this analysis highlights the potential of M3D IC technology to address the area overheads associated with asynchronous circuits, while also improving other performance parameters. By enabling compact layouts and efficient interconnect structures, M3D integration can pave the way towards making asynchronous logic more viable for future energy-efficient computing systems.

## V. Conclusions and Future Work

In this study, we integrate M3D technology with QDI Null Convention Logic (NCL) and propose a design methodology for the implementation of M3D-based NCL standard cells, aimed at mitigating the area inefficiencies of traditional planar or 2D counterparts. First, we designed M3D-NCL threshold cells and analyzed those in terms of crucial design parameters, such as delay, skew, power dissipation, and area utilization, followed by analytically comparing those to the corresponding 2D implementations. Simulation results suggest that, for a conservative wirelength reduction resulting from M3D imple-mentation, a substantial area reduction of 44% on average can be achieved while simultaneously reducing the average delay, skew, and power of the NCL gates by 11%, 10%, and 10%, respectively. The benefit of M3D-based implementations for large-scale NCL circuits was also studied, analyzed, and presented in this work. For an unsigned 4x4 NCL array multiplier, the M3D design outperformed the 2D version in all categories while utilizing half the area. In the future, we intend to develop an automated framework to create NCL standard cell layouts, leveraging existing CAD tools. We also plan to extend the work for other QDI paradigms such as Pre-Charge Half Buffers (PCHB) and Sleep Convention Logic (SCL).

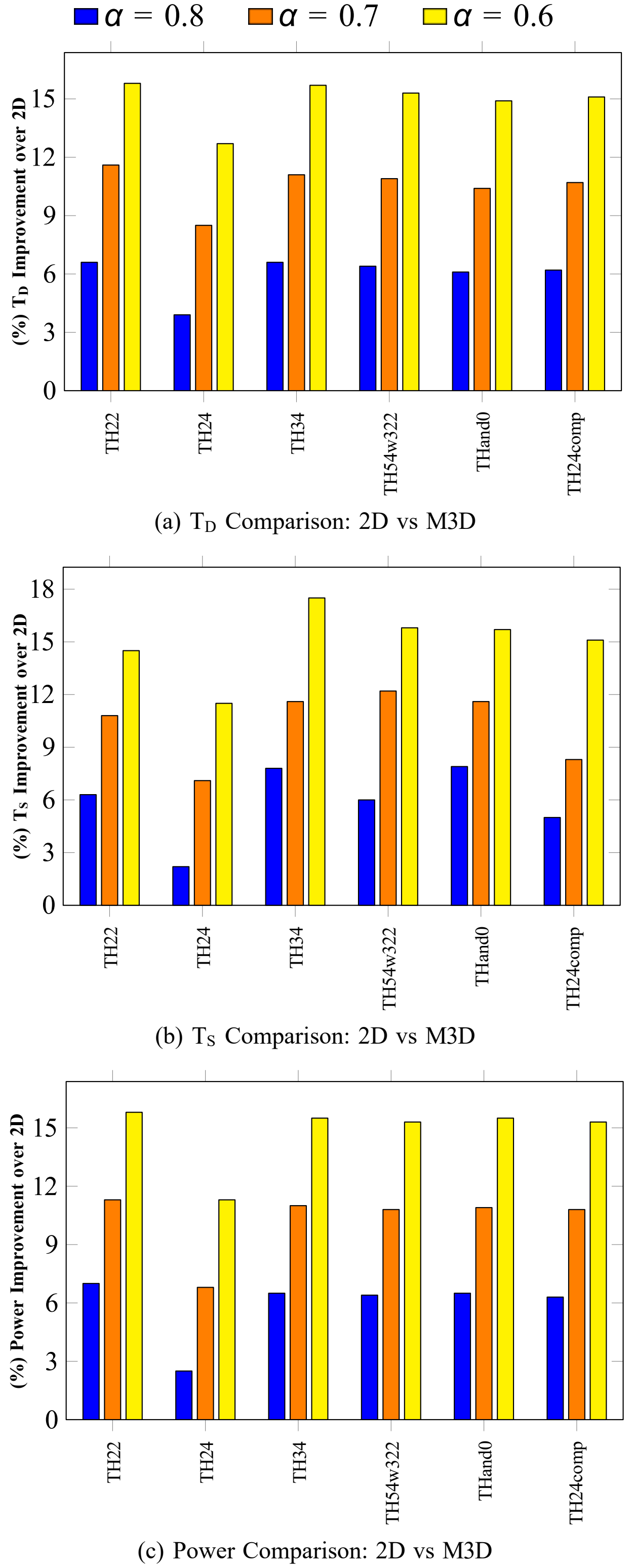

(a) $T_D$ Comparison: 2D vs M3D

(b) $T_S$ Comparison: 2D vs M3D

(c) Power Comparison: 2D vs M3D

Fig. 5: Power and delay improvements of M3D over 2D implementations with varying reduction factors ($\alpha$)

TABLE II: Conventional 2D vs M3D implementations of static NCL threshold gates ($\alpha = 0.7$)

| Gate | Case | $T_D$ (ps) | $T_S$ (ps) | Power (μW) | Area (μm²) |
|---|---|---|---|---|---|
| **TH22** | 2D | 96.4 | 69.6 | 1.74 | 0.2052 |
| | M3D | 85.2 | 62.1 | 1.55 | 0.1226 |
| | **% Improvement** | **11.6%** | **10.8%** | **11.3%** | **43.9%** |
| **TH24** | 2D | 154 | 95.7 | 2.07 | 0.4446 |
| | M3D | 141 | 88.9 | 1.93 | 0.2598 |
| | **% Improvement** | **8.5%** | **7.1%** | **6.8%** | **46.1%** |
| **TH34** | 2D | 208 | 124 | 2.19 | 0.4104 |
| | M3D | 185 | 110 | 1.95 | 0.2598 |
| | **% Improvement** | **11.1%** | **11.6%** | **11%** | **41.6%** |
| **TH54w322** | 2D | 164 | 110 | 2.00 | 0.3591 |
| | M3D | 146 | 96.7 | 1.79 | 0.2206 |
| | **% Improvement** | **10.9%** | **12.2%** | **10.8%** | **42.7%** |
| **THand0** | 2D | 158 | 105 | 1.98 | 0.3420 |
| | M3D | 141 | 92.8 | 1.77 | 0.2010 |
| | **% Improvement** | **10.4%** | **11.6%** | **10.9%** | **44.9%** |
| **TH24comp** | 2D | 153 | 102 | 2.01 | 0.3078 |
| | M3D | 137 | 93.7 | 1.79 | 0.1814 |
| | **% Improvement** | **10.7%** | **8.3%** | **10.8%** | **44.3%** |
| | **% Average** | **10.5%** | **10.2%** | **10.3%** | **43.9%** |

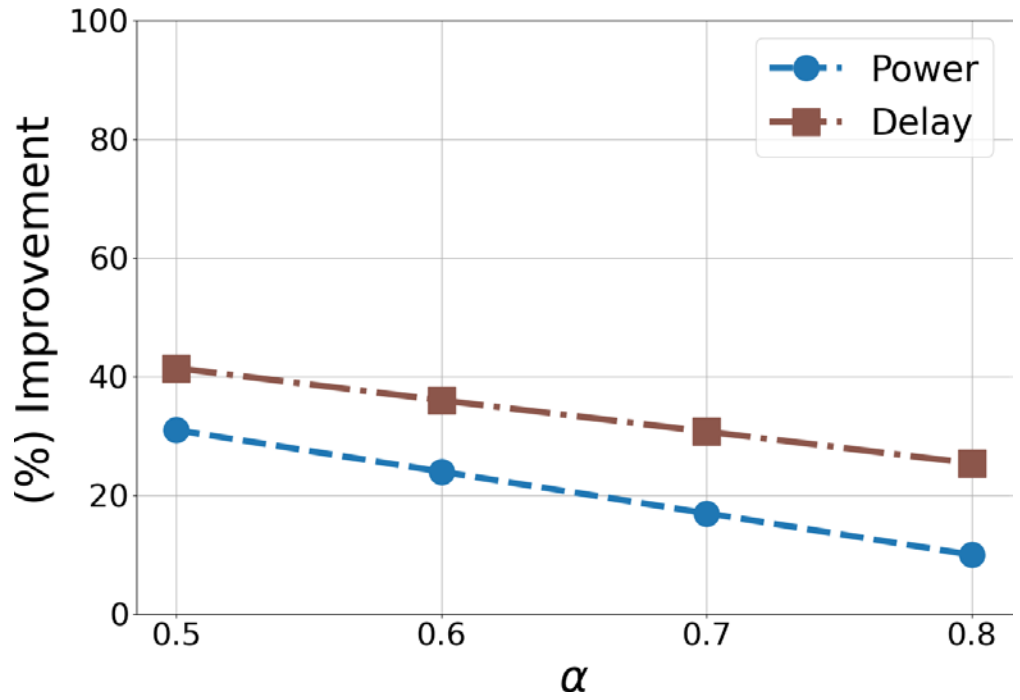

Fig. 6: (%) Improvement of M3D implementation over 2D for Multiplier circuit with varying $\alpha$